\input{aipcheck}

\documentclass[
    ,final            
  ]
  {aipproc}

\layoutstyle{6x9}

\begin{document}

\title{Color Transparency}

               
\author{Gerald A. Miller}{
  address={Physics Department, Univ. of Washington, Seattle, Wa. 98195-1560, USA}}

\begin{abstract}Color transparency is the vanishing of nuclear initial or final state interactions involving specific reactions.  The reasons for believing that color transparency might be  a natural consequence of QCD are reviewed. The main impetus for this talk is recent experimental progress, and this is  reviewed briefly.
  
\end{abstract}

\maketitle


\section{Introduction}

 This talk reviews the phenomena known as ``Color Transparency", including what it is, the experiments and the progress so far.  There have been several  reviews of color transparency~\cite{Frankfurt:1992dx,Frankfurt:1994hf,Jain:1995dd,Miller:2007zzd}. Much of the theory has been known for a long time, and it has not changed much since the time of those reviews. On the other hand,  there has been  significant  experimental progress in recent years, and one can anticipate much discovery work to be done in new  laboratories.
 
 The basic idea is that some times a hadron is in a color-neutral point-like configuration PLC.  If such undergoes a coherent reaction, in which one sums gluon emission amplitudes to calculate the scattering amplitude, the PLC does not interact with the surrounding media. A PLC is not absorbed by the nucleus. The nucleus casts no  shadow. This is a kind of quantum mechanical invisibility.
 
In more technical terms we speak of  reduced initial and//or  final state interactions in high momentum transfer 
quasi elastic nuclear reactions.  Examples include the electroproduction of a single meson in nuclear reactions,
and the coherent production of two jets by high energy pions incident on nuclei. 
The logic for color transparency to occur consists of three steps. 
\begin{itemize}
 \item high momentum transfer hadronic exclusive reactions proceed by PLC formation~\cite{ct}
\item  a PLC has a  small scattering amplitude because, for a color neutral object, the sum of gluon emission amplitudes cancel. A coherent process is needed.

\item a PLC expands as it moves~\cite{fs,jm}.
 Therefore  high energies are needed.
\end{itemize}

This means that to observe  color transparency one needs to make the PLC in a  high momentum transfer
 reaction using an  exclusive process to maintain the  coherence necessary for the cancellation of interactions. Furthermore,   high energy must be involved to avoid expansion of the PLC. In that case there are
 no or reduced interactions

Observing color transparency is interesting because this is a (relatively) new novel dynamical phenomenon in which the
strong interaction is turned off. Moreover, color transparency is routinely used  in QCD factorization proofs, so it is very worthwhile to  observe it experimentally. Furthermore, the existence of PLC holds  are many implications for nuclear 
physics~\cite{Frankfurt:1985cv,Frank:1995pv}.
In nuclei the PLC of the nucleons do not interact with the surrounding nucleons and this leads to modification of the nucleon structure, an effect that may be relevant for explaining the EMC effect.

The remainder of this talk is concerned with a 
 brief discussion of  the three main   points: dominance of PLCs  in high momentum transfer exclusive processes,  color cancellation which reduces or cancels the interactions of the PLC, and the irritating expansion effects. This is followed by 
 a very brief mentions of  
early searches using (p,pp)~\cite{Carroll:1988rp}
 and  (e,e',p)~\cite{O'Neill:1994mg} nuclear reactions. Then the case that worked- Fermilab di-jet experiment in coherent 
 pion nuclear reactions- is discussed~\cite{Frankfurt:1993it,dannyref}. This is followed by a discussion of the 
new results from Jefferson Lab involving electroproduction of pions~\cite{Clasie:2007aa}
and rho mesons~ \cite{ElFassi:2012nr}. A brief discussion of 
other possible experiments is presented.

\section{Three elements of Color Transparency}

\subsection{Small objects are produced at high momentum transfer in two-body scattering}
High momentum transfer is usually associated with small wavelengths. Perturbative QCD make the stronger assertion that a small object without soft color and pion fields is produced during a high momentum transfer. Such configurations of closely separated quarks can be produced with the fewest possible gluon exchanges. The use of perturbative QCD for exclusive reactions, at all but asymptotically large momentum transfers, has been questioned, but the inclusion of Sudakov effects extends the region of applicability. Moreover, theoretical analyses of the simplest versions of popular models show that small configurations can be produced for momentum transfers as low as about  1 or 2 (GeV/c)$^2$~\cite{Frankfurt:1992dx} . 

The question of whether high momentum transfer exclusive reactions requires  a PLC is interesting because there is an opposing idea known as the Feynman mechanism. In this case the process is dominated by the interactions of a single quark carrying nearly all of the momentum of the hadron, so that  the transverse size is not affected. Then no PLC is formed. The question of the  relevance of the PLC is the interesting question we want to learn about.

\subsection{Small objects have small cross sections}

This can be understood using     Coulomb interactions. Consider a dipole made of an electron-positron pair with a fixed  transverse separation  $b$ that interacts with a fixed charge $q$. The total Coulomb interaction $V=eq/r_+-eq/r_-\approx eq b/(r_+r_-)$, where $r_{+,-}$ is the distance between the (positron, electron) and the charge $q$, and one takes the longitudinal position of the pair to be at the target. The interaction is proportional to the dipole moment of the pair and vanishes when the transverse separation vanishes. Notice that we add the two interactions. This is where coherence enters. The strong interaction version of charge cancellation is used in proofs of factorization~\cite{Collins:1996fb}.

In QCD one deals with initial and final color singlet configurations. Thus two gluons must be exchanged and one gets an interaction proportional to $b^2$. This was originally derived for two  gluon exchange, but this  mechanism has difficulties. The slope of $d\sigma/dt $ diverges for the exchange of massless gluons. Including the effects of gluon-gluon interactions and the non-perturbative nature  of gluon-target scattering leads to the result~\cite{Blaettel:1993rd}:
\begin{equation}
\sigma(b^2)={\pi^2\over3} \left(b^2 \alpha_s(Q^2_{\rm eff}) x g_T(x,Q^2_{\rm eff})\right)_{x=\lambda/(sb^2)Q^2_{\rm efff}=\lambda/b^2},
\end{equation}
where $\lambda$ is a process-dependent proportionality factor.

\subsection{PLC Expansion}
The point-like configuration is not an eigenstate of the Hamiltonian. It can be regarded as a wave packet that undergoes time evolution. Since the PLC (by definition) starts out as small-sized, it can only increase in size. Ultimately it will turn into a normal-sized hadron that interacts with the usual strength. Therefore the observation of color transparency requires that the PLC escape the nucleus before it expands.  For a protonic PLC of momentum $P$, the expansion time $t \approx 1/(M_n-M_p) (2P/(M_n+M_p)=2P/\Delta M^2$, where $M_p$ is the proton mass and $M_n$ is the mass of an important component of the PLC. 
The time $t$ can be thought of as the time-dilated version of a natural rest-frame time. For large enough values of $P$ the 
PLC remains small throughout its transversal of the nucleus.

\begin{figure}\label{exp}
  \includegraphics[height=.25\textheight]{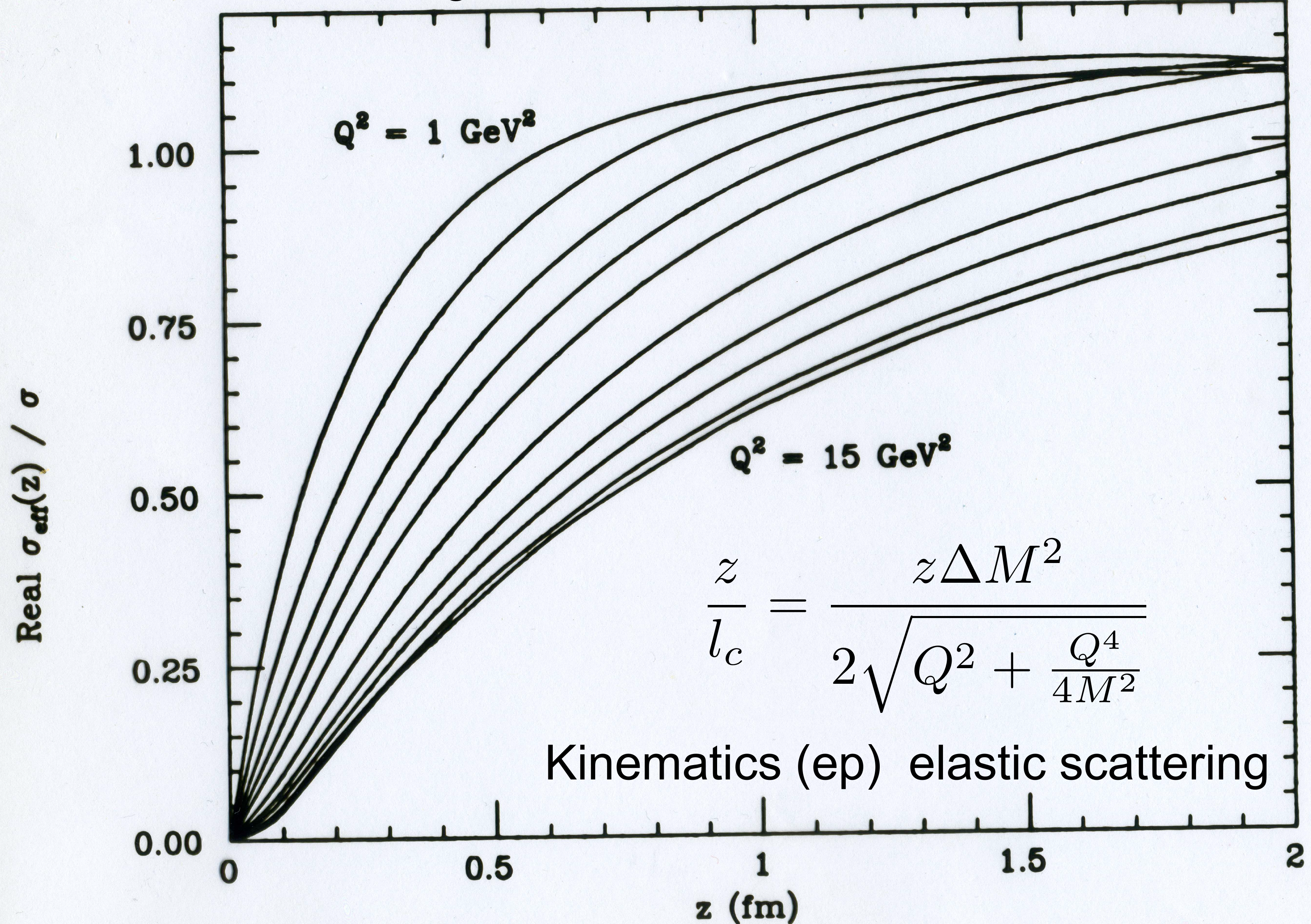}
  \caption{Effective cross section as a function of the distance $z$ transversed, from PRL 69, 3619\cite{jm}}
\end{figure}
Jennings \& Miller~\cite{jm}  used  a hadronic basis to understand the expansion. Their results for the effective cross section as a function of distance transversed $z$  are displayed  in Fig.~1. This function is rather similar to the 
results of the diffusion model of Strikman \& Frankfurt~\cite{fs}:
\begin{equation}
\sigma_{\rm eff}(z,P)=\sigma(P)\left[\left({n^2\langle k_t^2\rangle\over Q^2}(1-{z\over l_c})\right)\theta(l_c-z)+\theta(z-l_c)\right],\;l_c\equiv 2P/\Delta M^2,
\end{equation}
where $\sigma$ is the ordinary hadronic cross section, $n$ the number of constituents in the PLC, a reasonable numerical value for $\Delta M^2=0.7$ GeV$^2$.

\section{Experimental consequences}
The   transparency $T(A)$ for a nucleus $A$ is defined as the ratio of a measured cross section to one computed in the absence of final state interactions. In the limit of complete transparency $T(A)=1$. At low momentum transfer $Q^2$
 $T(A)$ takes on a value that is characterized by the ordinary distorted wave Born approximation (DWBA).  But as the value of  $Q^2$  increases $T(A)$ increases, assuming that the associated  momentum $P$ increases with increasing $Q^2$.
 An generic illustration is shown in Fig.~2.  Since the axes are not labelled with numbers and the scale is unknown, the 
 curve refers to all experiments and all models for computing $T(A)$.
 \begin{figure}\label{ta}
  \includegraphics[height=.225\textheight]{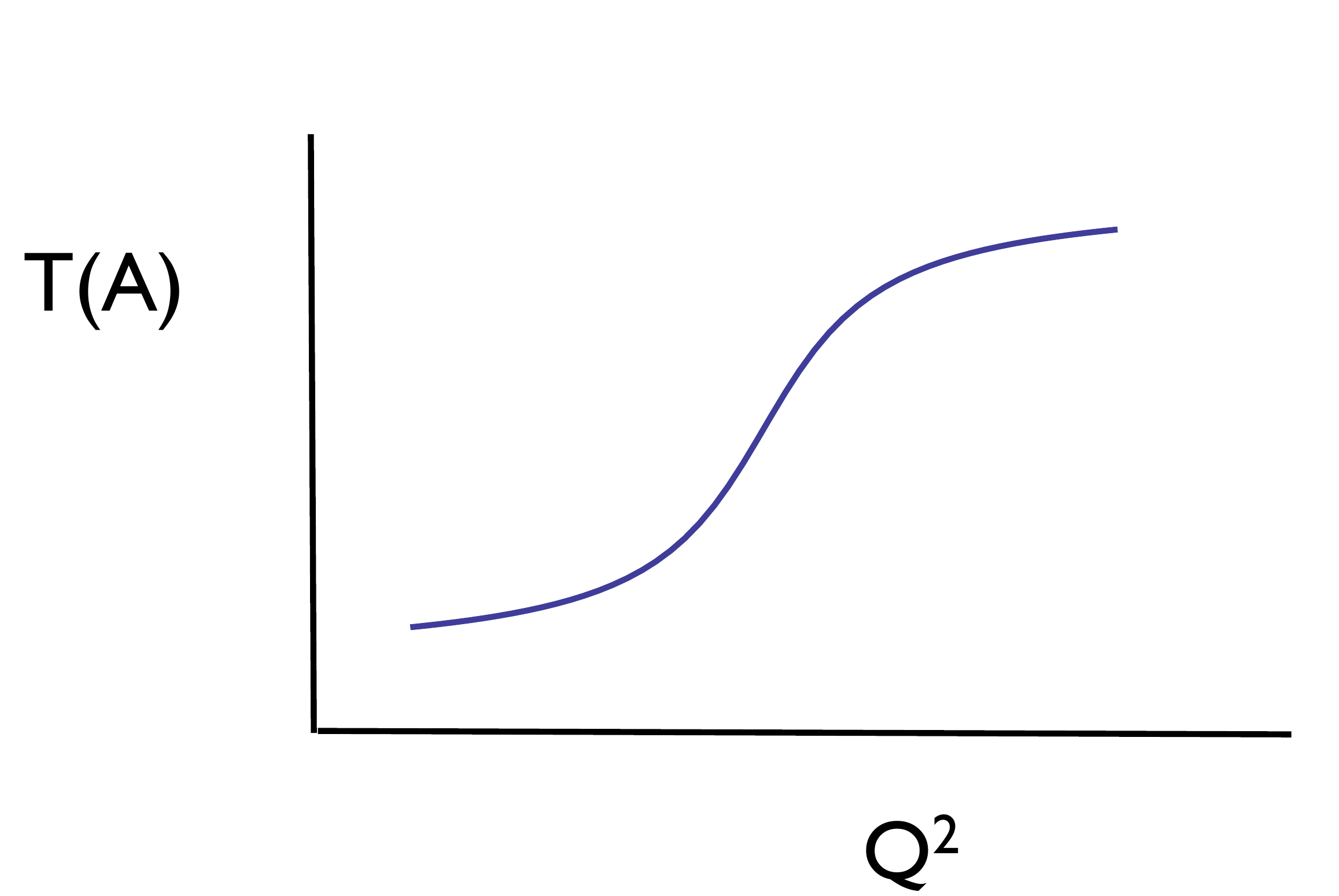}
  \caption{Generic transparency as a function of $A$.}
\end{figure}

The remaining discussion is brief; please see the talk~\url{https://indico.triumf.ca/confAuthorIndex.py?view=full&letter=m&confId=1383} for more information.The BNL (p,pp) experiment~\cite{Carroll:1988rp}, which showed a rise and then a fall of $T(A) $ with
increasing beam momentum (at fixed cm pp scattering angle). This finding   has not  been 
reproduced by any theory that included all of the necessary inputs: expansion, effects of PLC and effects of large blob-like configurations, BLC.

The (e,e'p) reaction shows no effects of color transparency~\cite{O'Neill:1994mg}, but future running at JLab12 could find an effect.

The coherent production of di-jets from nuclei using the 500 GeV FermiLab pion beam~\cite{dannyref} is the one that worked.
It shows color transparency through Pt to C target ratio that is seven times larger than that predicted from Glauber theory~{Frankfurt:1993it}.
The pion is a likely object for color transparency involvement because it has a singular central transverse charge density~\cite{Miller:2009qu}.

There are encouraging signs of color transparency in the JLab $ (e,e',\pi^+)$ experiment~\cite{Clasie:2007aa}.   A significant rise of $T(A)$ as 
$Q^2$ ranges between 1 and 5 (GeV/c)$^2$ has been detected. More running is planned at JLab12.

The $(e,e',\rho^0)$ experiment\cite{ElFassi:2012nr}, also at JLab, has detected a significant rise of $T(A)$ as $Q^2$ ranges between  
1 and 2.5 (GeV/c)$^2$.  More running is planned at JLab12.

\section{Summary}
Color transparency is an expected, but not certain, consequence of QCD. It has been observed at high energies
at FermiLab. Evidence at medium energy is piling up. It seems that PLC formation is an important part of 
(single) meson production at large values of $Q^2$, but has not yet been observed for the nucleon.


\begin{theacknowledgments}
I thank the USDOE for partial support of this work. I thank my  colleagues on color transparency projects
L. Frankfurt, B. K. Jennings and  M. Strikman for  many efforts. 
\end{theacknowledgments}



\bibliographystyle{aipproc}  

\begin{thebibliography}{99}

\bibitem{Frankfurt:1992dx} 
  L.~Frankfurt, G.~A.~Miller and M.~Strikman,
  \emph{ Comments Nucl.\ Part.\ Phys.} {\textbf 21}, pp. 1--40  (1992).
\bibitem{Frankfurt:1994hf} 
  L.~L.~Frankfurt, G.~A.~Miller and M.~Strikman,
 \emph{   Ann.\ Rev.\ Nucl.\ Part.\ Sci.}  {\textbf 44}, pp. 501--560 (1994).

\bibitem{Jain:1995dd} 
  P.~Jain, B.~Pire and J.~P.~Ralston,
 \emph{   Phys.\ Rept. }  {\textbf 271}, pp.  67--179 (1996).
  
\bibitem{Miller:2007zzd} 
  G.~A.~Miller,
  ``Colour Transparency,'' in \emph{Electromagnetic interactions and hadronic structure}, edited by
  F. Close,   et al., Cambridge University Press,Cambridge, 2007,  pp. 457--494
 \bibitem{ct}  A.H. Mueller, in \emph{Proceedings of Seventeenth Rencontre de
    Moriond, Les Arcs} edited by J Tran Thanh Van, Editions Frontieres,
Gif-sur-Yvette, France, 1982, Vol. I, pp. 13--43;
S.J.~Brodsky in \emph{Proceedings of the Thirteenth Int'l Symposium
on Multiparticle Dynamics}, edited by W. Kittel, W. Metzger and A. Stergiou
World Scientific, Singapore 1982, pp. 963--1002.
\bibitem{fs} L. Frankfurt and M. Strikman,  \emph{Phys. Rep.} {\bf 160}, pp.
235--427, (1988).
\bibitem{jm} B.K. Jennings and G.A. Miller,  \emph{Phys. Lett. B} {\bf 236},
pp. 209--213 (1990);  
B.K. Jennings and G.A. Miller,  \emph{Phys. Rev. D} {\bf 44},
pp. 692--703 (1991); 
 \emph{Phys. Rev. Lett.} {\bf 69}, pp. 3619--3622 (1992); 
 \emph{Phys. Lett.B }
{\bf 274}, pp. 442--448 (1992).

\bibitem{Frankfurt:1985cv} 
  L.~L.~Frankfurt and M.~I.~Strikman,
 \emph{ Nucl.\ Phys.\ B} {\bf 250}, pp. 143---176 (1985).
\bibitem{Frank:1995pv} 
  M.~R.~Frank, B.~K.~Jennings and G.~A.~Miller,
\emph{  Phys.\ Rev.\ C} {\bf 54}, pp. 920--935 (1996)
\bibitem{Carroll:1988rp} 
  A.~S.~Carroll,   {\it et al.},
  \emph{ Phys.\ Rev.\ Lett.}  {\bf 61}, 1698--1701 (1988);
  J.~Aclander,   {\it et al.},
  \emph{  Phys.\ Rev.\ C} {\bf 70}, 015208 (2004).

\bibitem{O'Neill:1994mg} 
  T.~G.~O'Neill  {\it et al.},
 \emph{   Phys.\ Lett.\ B} {\bf 351}, pp. 87--92 (1995).;
  N.~Makins,  {\it et al.},
  \emph{  Phys.\ Rev.\ Lett.}\  {\bf 72}, 1986--1989 (1994);
 K.~Garrow,  {\it et al.},
 \emph{   Phys.\ Rev.\ C} {\bf 66}, 044613 (2002)

\bibitem{Frankfurt:1993it}
  L.~Frankfurt, G.~A.~Miller and M.~Strikman,
 {\emph Phys.\ Lett.\ }   {\bf B304}, pp. 1--7 (1993); L.~Frankfurt, G.~A.~Miller and M.~Strikman,
{\emph  Phys.\ Rev.\ D} {\bf 65}, 094015 (2002);
 L.~Frankfurt, G.~A.~Miller and M.~Strikman,
{\emph   Found.\ Phys.}\  {\bf 30}, pp.  533--542 (2000).
\bibitem{dannyref} 
  E.~M.~Aitala {\it et al.}  [E791 Collaboration],
  Phys.\ Rev.\ Lett.\  {\bf 86}, pp. 4773-- 4777(2001).
\bibitem{Clasie:2007aa} 
  B.~Clasie,   {\it et al.},
 \emph{   Phys.\ Rev.\ Lett.}\  {\bf 99}, 242502 (2007) 
 \bibitem{ElFassi:2012nr} 
  L.~El Fassi,   {\it et al.},
\emph{ Phys.\ Lett.\ B} {\bf 712}, pp. 326---330 (2012).
  

\bibitem{Collins:1996fb} 
  J.~C.~Collins, L.~Frankfurt and M.~Strikman,
 \emph{   Phys.\ Rev.\ D} {\bf 56}, pp. 2982--3006 (1997).
  \bibitem{Blaettel:1993rd} 
  B.~Blaettel, G.~Baym, L.~L.~Frankfurt and M.~Strikman,
   \emph{ Phys.\ Rev.\ Lett.}\  {\bf 70}, pp. 896--899 (1993).
\bibitem{Miller:2009qu} 
  G.~A.~Miller,
   \emph{ Phys.\ Rev.\ C }{\bf 79}, 055204 (2009);
   G.~A.~Miller, M.~Strikman and C.~Weiss,
 \emph{   Phys.\ Rev.\ D} {\bf 83}, 013006 (2011)
\end{thebibliography}

\end{document}